\documentclass[runningheads]{llncs}
\usepackage[T1]{fontenc}
\usepackage{graphicx}
\begin{document}
\title{User Experience In Dataset Search}
%
%
\author{Yihang Zhao\inst{1}\orcidID{0009-0009-2436-8145} \and
Albert Meroño Peñuela\inst{1}\orcidID{0000-0003-4646-5842} \and
Elena Simperl\inst{1}\orcidID{0000-0003-1722-947X}}
\authorrunning{Y. Zhao et al.}
%
\institute{King's College London\\
\email{\{yihang.zhao,albert.merono,elena.simperl\}@kcl.ac.uk}}
\maketitle              
\begin{abstract}
This research investigates User Experience (UX) issues in dataset search, targeting Google Dataset Search and data.europa.eu. It focuses on 6 areas within UX: Initial Interaction, Search Process, Dataset Exploration, Filtering and Sorting, Dataset Actions, and Assistance and Feedback. The evaluation method combines 'The Pandemic Puzzle' user task, think-aloud methods, and demographic and post-task questionnaires. 29 strengths and 63 weaknesses were collected from 19 participants involved in roles within technology firm or academia. While certain insights are specific to particular platforms, most are derived from features commonly observed in dataset search platforms across a variety of fields, implying that our findings are broadly applicable. Observations from commonly found features in dataset search platforms across various fields have led to the development of 10 new design prototypes. Unlike literature retrieval, dataset retrieval involves a significant focus on metadata accessibility and quality, each element of which can impact decision-making. To address issues like reading fatigue from metadata presentation, inefficient methods for results searching, filtering, and selection, along with other unresolved user-centric issues on current platforms. These prototypes concentrate on enhancing metadata-related features. They include a redesigned homepage, an improved search bar, better sorting options, an enhanced search result display, a metadata comparison tool, and a navigation guide. Our aim is to improve usability for a wide range of users, including both developers and researchers.

\end{abstract}
\section{Introduction}
In the present data-driven landscape, sectors such as healthcare and social sciences are increasingly reliant on diverse datasets \cite{benjelloun2020google,mayer2013big}. The sheer volume and complexity of available datasets often result in data silos, hampering their effective usage \cite{roh2019survey,stonebraker2018data}. This dichotomy between the surfeit of publicly available datasets and the particular datasets users can actually locate constitutes a significant challenge \cite{chapman2020dataset}.

Earlier studies have delved into the intricacies of dataset search systems, highlighting the critical role of metadata and the need for its standardisation \cite{chapman2020dataset,benjelloun2020google}. Research has also been conducted on how Google Dataset Search leverages metadata to organise and rank search results \cite{brickley2019google}. Additionally, some focus has been given to understanding user behaviour on governmental dataset platforms, identifying the primary factors that influence search queries \cite{kacprzak2019characterising}. However, there still exists a void in understanding the user experience in navigating these dataset search platforms, an area this research aims to explore.

For this investigation, Google Dataset Search and data.europa.eu were selected due to their advanced metadata-based search features and high prevalence among researchers. This study assesses user experience in both document retrieval and dataset retrieval, examining existing features of Google Dataset Search and data.europa.eu platforms. Researchers synthesis these elements to pinpoint six aspects of user experience in dataset search. A user task, termed "The Pandemic Puzzle," is then designed to explore currently trending areas in dataset search. This study adopts both Concurrent and Stimulated Retrospective think-aloud techniques to investigate user interactions with these dataset retrieval platforms. To enhance the reliability of the think-aloud methods, Thought Prompts and preliminary training sessions are included. Additional context and a swift assessment of perceived usability are provided through demographic details and post-task questionnaires.

This research contributes to enhancing user experience on dataset search platforms in several ways. Firstly, the redesigned homepage emphasizes the platform's scope, clearly showing content that users are most interested in, aiding them in determining if the site meets their data needs. Secondly, the search bar extends beyond datasets to include non-dataset items like help documents and authors, while providing visual aids for Boolean operators. These operators are effective for leveraging metadata, significantly improve the efficiency of finding relevant datasets. Thirdly, the dataset preview panel using color-blind-friendly colors to represent metadata quality and displaying the most sought-after metadata to reduce reading fatigue and hasten dataset filtering. Fourthly, simplified filters with a display of the remaining dataset count prevent over-filtering. Fifthly, a metadata comparison feature utilizes colors to differentiate quality between two datasets on the same metadata, speeding up the final selection process. Lastly, a visual tutorial clarifies the user interface functions, offering guidance for new users.

These research findings serve as a basis for improving dataset search platforms. The results guide platform developers in understanding user behavior and preferences, aiding in the creation of more efficient dataset discovery tools.
\section{Related Works}
\label{sec:related_work}

\textbf{Information seeking behaviour:} in web searches and digital libraries, information seeking behaviour is a well-studied area with established models that map out the process and strategies users follow ~\cite{ellis1989behavioural}. However, the suitability of these models for dataset retrieval is uncertain since they were originally intended for document searches ~\cite{ellis1989behavioural}. Searches are generally of two types: 'lookup', which is simple and well-supported, and 'exploratory', which is more complex, involving tasks that require users to learn and discover new information ~\cite{marchionini2006exploratory}. Exploratory searches can be open-ended, leading to uncertainty when information is incomplete ~\cite{marchionini2006exploratory}. The complexity of exploratory searches is especially significant in dataset retrieval, which has become more challenging with initiatives like GO FAIR that encourage the sharing and reuse of datasets. Such initiatives have led to a greater volume and variety of data, highlighting the need for dataset search systems to be designed with the user's needs in mind, helping them navigate the complex process of finding the right datasets ~\cite{kern2015there}.

\textbf{Nature of datasets:} Chapman et al. ~\cite{chapman2020dataset} describe a dataset as an ordered collection of either numerical or textual data points. Metadata, crucial for search enhancement and contextualisation, includes elements like data source, format, and collection time. Their work identifies two core dataset search approaches: 'basic' and 'user-organised.' The 'basic' search is found on platforms such as Figshare~\cite{thelwall2016figshare} and Google Dataset Search~\cite{brickley2019google}, where datasets are pre-curated and stored in repositories. The 'user-organised' search takes place in environments like data lakes~\cite{gao2018navigating} and data markets~\cite{grubenmann2018financing}, where users assemble individual data points themselves, as seen in climate studies collecting temperature and humidity metrics in a data lake.

\textbf{Online Dataset Accessibility:} Benjelloun et al.~\cite{benjelloun2020google} examined Schema.org ~\cite{schemaorg2024}, noted for having the largest and most varied corpus of its kind, and found that only 34\% of datasets include licensing information, while 44\% offer data download links. Additionally, metadata coverage generally falls under 50\%, affecting dataset reusability and findability, thereby influencing user experience in dataset searches. Their study also highlights that 37\% of datasets lack persistent URLs, underscoring the necessity for stable identifiers like DOIs. Their findings advocate for standardisation in dataset descriptions, licensing, and access. While transitional periods will be required for standardisation, immediate improvements in user experience are achievable.

\textbf{Mechanism for searching datasets:} Brickley et al.~\cite{brickley2019google} describe the dataset search engine's multi-stage process, starting with web crawlers collecting metadata across the web. This metadata is subsequently standardised and structured into a universal format using frameworks like Schema.org. Cleaned metadata is then matched with a pre-existing database, or a knowledge graph, to add contextual richness to each dataset. The engine assigns unique 'fingerprints' to datasets to group similar datasets and eliminate redundancies. Brickley et al. also discuss the engine's use of MapReduce~\cite{brickley2019google} for efficient data processing. Daily scans are performed to capture new or updated datasets. During a search, the engine ranks datasets based on their query relevance, continually updating its methods to improve its effectiveness and reach.

\textbf{User-centric literature retrieval:} from literature ~\cite{ghorab2013personalised,kelly2015development,croft2010search}, they summary 10 user considerations: 'Relevance of Results' focuses on query interpretation accuracy, context sensitivity, and result ordering; 'Ease of Use' highlights intuitive interfaces and straightforward navigation; 'Performance and Speed' aims for quick responses and system reliability; 'Search Features' includes advanced search options and auto-suggestions; 'Quality of Information' stresses source credibility and data precision; 'Security and Privacy' emphasizes user consent and encryption; 'Adaptability' caters to different languages, regions, and devices; 'Presentation of Results' targets clear labelling and data visualisation options; and 'Personalisation' utilises user history and preferences for customisation. In addition, 'Accessibility features', such as voice search and readability, are integrated to ensure inclusive design.

\textbf{User-centric dataset retrieval:} from literature ~\cite{chapman2020dataset,kacprzak2019characterising,benjelloun2020google}, they indicates that factors like search result accuracy, user interface clarity, search speed, data protection, and system adaptability are crucial . Four key areas for tool enhancement emerge: 'Dataset Metadata' should have comprehensive descriptions, standardisation, frequent updates, and be optimised for search; 'Functionality, Support, \& Security' implies powerful search features, reliable support, robust data security, and wide accessibility; 'Usability' entails a user-friendly interface, fast performance, and effective visualisation; 'Support' is augmented by detailed documentation, educational resources, active user communities, and ongoing tool improvements.

\textbf{Comparing dataset retrieval with traditional literature retrieval:} a study ~\cite{kern2015there} with 53 participants highlighted the necessity for dataset retrieval systems to be tailored to the specific needs of researchers, as their decision-making process is more complex than that for literature retrieval. The study's key findings emphasized the significance of metadata quality and accessibility, with users showing a willingness to contribute to metadata enhancement. Additionally, the research found that users value supplementary literature for added context and prefer search systems that are intuitive, like those with autocomplete features, and that offer research-relevant updates and suggestions.

\textbf{User study on dataset search platform:} Kacprzak et al.~\cite{kacprzak2019characterising} scrutinise user interactions on government dataset search platforms, employing both qualitative and quantitative methods. They pinpoint two significant factors influencing user queries: data features and structural layout. Their research indicates user queries often include geospatial and temporal elements. Metrics reveal that search activity peaks during typical working hours, likely indicating professional use. Their study's data also show most users access platforms via search engines and predominantly use browsers like Chrome, although Internet Explorer retains significant usage, possibly due to its corporate prevalence. They observe that user engagement metrics, such as page views and time spent, increase with return visits, and they identify challenges users face in data discovery as a prospective area for enhancing search functionalities.

\section{Approach}
The entire section 3 unfolds following the sequential order of the project's design process. To establish targeted user tasks, researchers first examined all features of the two websites. They then integrated details from user-centered literature and dataset retrieval in the related works to design the final user tasks. Subsequently, corresponding user study methods and recruitment strategies were set up, during which solutions to challenges were proposed.
\subsection{Explore Platform Features}
\label{appro:features}
Researcher reviews 2 dataset search platforms to document their features and evaluate their functionalities, aiming to create a task for assessing user experience.
\subsubsection{Google Dataset Search}
\label{appro:google_features}
\begin{figure}[h]
    \centering
    \includegraphics[width=0.8\linewidth]{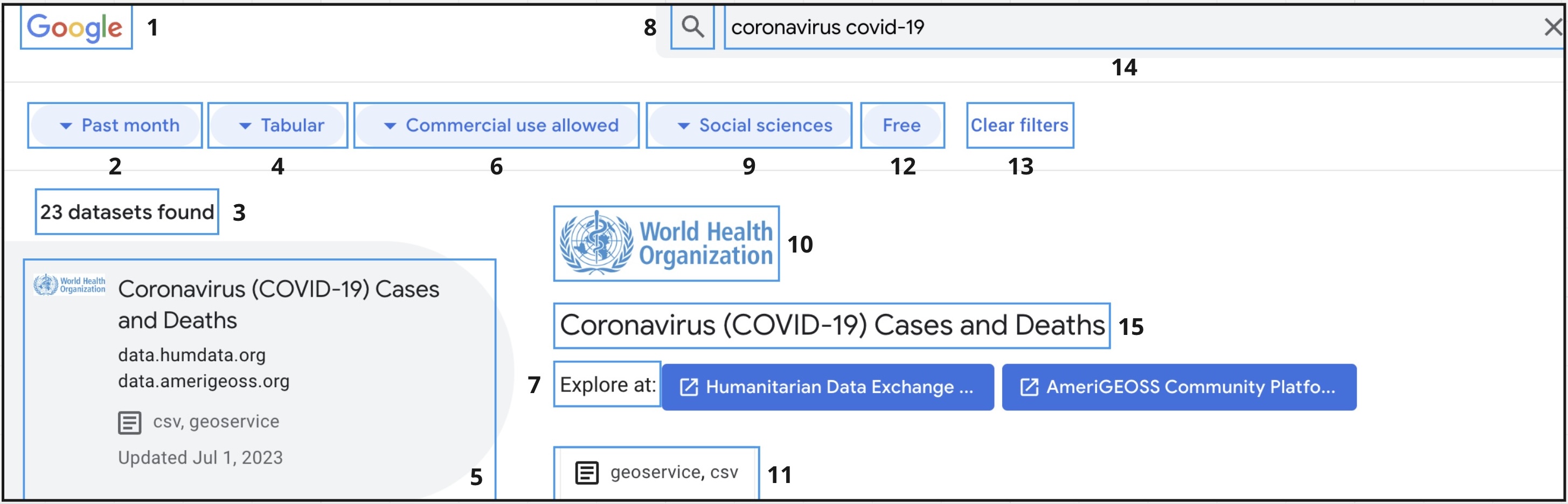}
    \caption{Google Dataset Search Platform Features}
    \label{fig:google_platform_features}
\end{figure}

Offers 38 core features intended to streamline user interaction and assist data acquisition. Features such as filters for file type, update time, usage rights, and subject categories enable users to personalise search results. Additional functionalities like dataset saving and sharing foster collaborative research. Dataset citation counts are also displayed as a relevance metric. Detailed file information and dataset previews facilitate quick initial evaluations. Query input is enhanced by autocomplete suggestions. Metadata attributes, including identifiers and author details, are visibly displayed. Figure~\ref{fig:google_platform_features} illustrates 15 primary features on the main page.
\subsubsection{data.europa.eu}
\label{appro:eu_features}
\begin{figure}[h]
    \centering
    \includegraphics[width=0.8\linewidth]{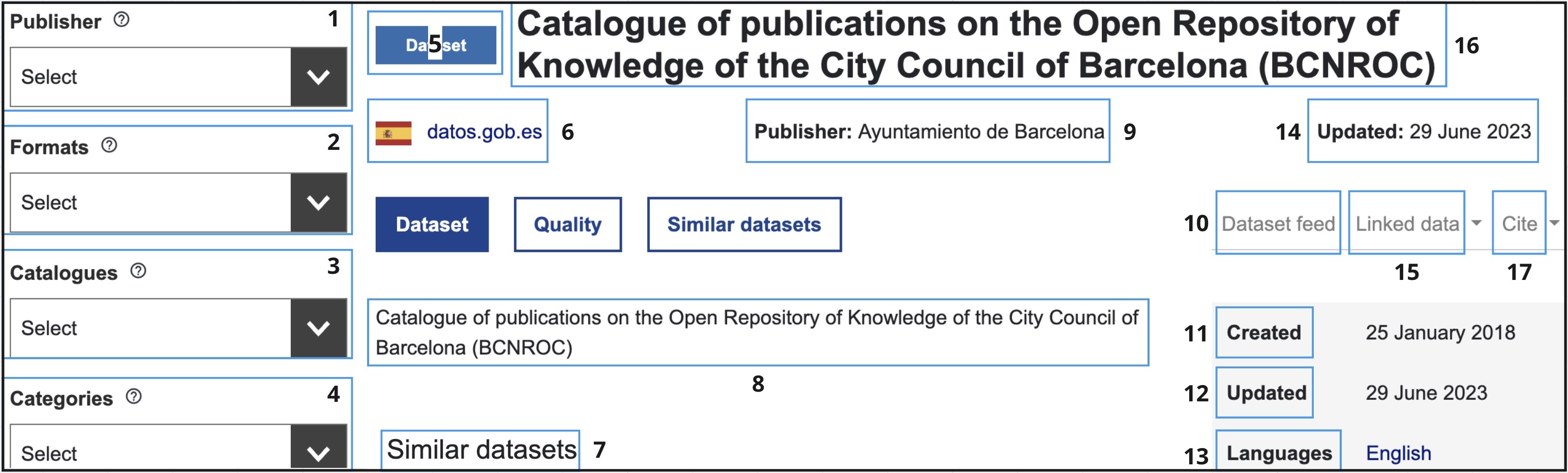}
    \caption{data.europa.eu Platform Features}
    \label{fig:eu_platform_features}
\end{figure}

Data.europa.eu expands on Google Dataset Search by offering advanced filtering options, diverse search types, and custom sorting. It allows users to view recent updates, multiple download options, and comprehensive metadata. Quality indicators and similar dataset suggestions are available. User navigation is monitored, and a SPARQL search enables complex queries. Users can view and visualise datasets directly without downloading. Figure~\ref{fig:eu_platform_features} illustrates 17 primary features on the main page.

\subsection{Explore User Experience Aspects}
Following an analysis of both user requirements and platform functionalities, this study delineates six facets of User Experience: Initial Interaction scrutinises the landing page elements like search buttons and language settings for ease of navigation. Search Process evaluates the efficiency of the search mechanisms, considering factors like speed, accuracy, and result relevance. Dataset Exploration focuses on the user-friendliness of accessing individual dataset details, including readability and source information. Usage of Filters and Sorting assesses the relevance and ease of use for sorting and filtering options. Dataset Actions looks at how conveniently users can perform actions like saving and sharing datasets. Finally, Feedback and Help analyses the platform's support features, emphasising user ease in providing feedback and seeking help.

\subsection{Design User Task}
This task in Table ~\ref{tab:practice_task_1} coined 'The Pandemic Puzzle' serves as a mechanism to understand user interaction in identifying, assessing, and utilising datasets on COVID-19 and social sciences.

\begin{table}
\caption{The Pandemic Puzzle User Task}
\label{tab:practice_task_1}
\begin{scriptsize}
\begin{tabular}{|p{0.2\columnwidth}|p{0.78\columnwidth}|}
\hline
\textbf{User Task} & \textbf{Instructions and Considerations} \\ \hline
Background & As a health sciences researcher, investigate COVID-19's impacts on the UK's life and social sciences sectors. Focus on datasets that are current, freely accessible, commercial-use, well-documented, from reputable sources, and specific to UK regions in tabular form. \\ \hline
1. Initial Interaction & Engage with the platform to locate datasets on 'COVID-19 impact on UK healthcare systems' and 'social science research during pandemics'. Assess interface usability including visibility of search, delete, and return functions. \\ \hline
2. Search Process & Input specific queries like 'COVID-19 mental health data UK' and observe platform's auto-complete suggestions, result speed, and relevance to the UK's life and social sciences. \\ \hline
3. Dataset Exploration & Scrutinize datasets for relevance, examining names, sources, types, update frequency, geographic specificity, and alignment with the research on UK's response to COVID-19. \\ \hline
4. Filters and Sorting & Utilize filters to narrow results to recent, open-access datasets, applying sorting to prioritize by relevance or recency. Assess the precision of filtering and sorting mechanisms. \\ \hline
5. Dataset Actions & Identify suitable datasets for longitudinal analysis of the pandemic's effects, noting options for saving and sharing. Evaluate the efficiency and intuitiveness of these processes. \\ \hline
6. Feedback and Help & Document any usability issues encountered and identify how the platform provides assistance. Explore the process for submitting feedback and seeking help within the platform. \\ \hline
\end{tabular}
\end{scriptsize}
\end{table}

\subsection{Use of User Study Methods}
\subsubsection{Use of Concurrent Think-Aloud}
This study uses moderate-complexity tasks, guided by \cite{kelly2009methods}. Participants perform tasks that mix quick reactions with thoughtful analysis, such as choosing a relevant dataset and analysing its metadata. These tasks are more involved than basic queries but not as complex as detailed data analysis. This research employs a think-aloud technique as suggested by \cite{charters2003use}, giving real-time insight into participants' thinking. VooV software captures both audio and visual data for comprehensive documentation.

\subsubsection{Challenges in Concurrent Think-Aloud}
One issue is that participants may feel awkward vocalising their thoughts, which can affect natural decision-making and skew results. Participants might also forget to verbalise their thoughts, especially when focused or performing difficult tasks. Another challenge is the risk of cognitive overload due to multitasking, especially in tasks requiring intense concentration \cite{charters2003use}.

\subsubsection{Addressing Challenges During Concurrent Think-Aloud}
\paragraph{Pre-Study Training}
A practice task requires participants to find a credible COVID-19 vaccination article. This task has four steps: defining the search goal, conducting the search while verbalising, evaluating sources aloud, and choosing an article while explaining the reasons. A reflection stage follows where participants can identify what worked and what needs improvement.

\paragraph{Providing Thought Prompts}
Used to help participants articulate their thoughts during this study. These are specific questions or statements aimed at prompting real-time reactions as participants use dataset search platforms. The goal is to get a detailed understanding of the user experience. Examples of thought prompts include: "What do you think of this dataset after selecting it?" and "Which filters are you using and why?".
\subsubsection{Use of Stimulated Retrospective Think-Aloud}
This research incorporates both Stimulated Retrospective Think-Aloud (SRTA) and Concurrent Think-Aloud (CTA) techniques to achieve a comprehensive view of user interactions~\cite{sela2000comparison,charters2003use}. The SRTA method allows participants to reflect on their actions and thoughts, using visual aids or playback footage for reference. Structured SRTA questions guide this reflective process, creating a reliable setting for participants to share their experiences, thereby improving data quality. An example from this study illustrates this approach.
\begin{itemize}
\item \textbf{Retrospective Thought Prompt:} "Reflecting on the search, do you remember your initial reactions? Did the results meet your expectations?"
\item \textbf{Possible Retrospective Answer:} "Initially, I was shocked by the sheer volume of information. I now realise my search terms were too broad, leading to the overload. If I were to do it again, I'd use more specific search criteria to narrow down the results."
\end{itemize}
In the immediate response, the participant notes the unexpected volume of results. The retrospective answer goes further by identifying the reason for the initial surprise and suggesting a way to improve future searches.
\subsection{Implement Questionnaires}
\subsubsection{Demographic Questionnaires}
Utilised at commencement to gain insights into participant backgrounds~\cite{boynton2004selecting}. These questionnaires capture specifics such as professional skills, educational history, and prior familiarity with dataset search utilities. Comprising seven questions, they targets broad categories like 'General', 'Education', and 'Occupation'. It solicits information about age group, gender, highest educational attainment, work or study field, frequency of dataset search tool usage, skill level, and previously used tools. Both multiple-choice and open-ended question types are employed.
\subsubsection{Post-Task Questionnaires}
Issued upon task completion to collect user opinions~\cite{lewis2014usability}. They examine usability facets including user satisfaction, perceived utility, and suggested refinements. These surveys encompass four main areas: 'Initial Interaction', 'Search Operation', 'Dataset Study', and 'Filter and Sort Utilisation'. Questions cover a range of topics, from initial impressions and the relevance of search outcomes to the quality of dataset metadata and the efficacy of additional features. A 1-5 rating scale is used for certain queries, while open-ended questions solicit suggestions for enhancements.
\subsection{Participant Recruitment}
This study involved 19 participants from diverse backgrounds, including academia and industry, recruited through LinkedIn, Boss Zhiping, and university email campaigns. This group consisted of 2 UX designers, 2 product managers, 1 special education aide, and 14 students specializing in Human-Computer Interaction and health data science, with ages ranging from 18 to 40. Ethical approval was obtained, and informed consent was collected from all participants. All participants tested both tools using the same task.

This study used scenarios related to COVID-19 and social sciences to motivate user interaction with the platforms, focusing on overall user experience, behavior, functionality, and interface design across various search websites. The scenarios prompted users to test specific platform features, such as autocomplete responses to 'COVID-19' and filter effectiveness in the 'social sciences' category. The research did not aim to draw conclusions specific to social sciences, which is why participants were recruited from a wide range of fields.
\subsection{Transcription and Coding Scheme}
A single researcher manually transcribed 19 videos, organizing the data into 6 primary themes, including 'Initial Interaction' and 'Search Process'. The coding process combined top-down and bottom-up approaches as outlined by Gu et al.~\cite{gu2014code}. Preconceived notions about web search biases, such as prior beliefs ~\cite{white2014belief}, informed some of the top-down codes, whereas bottom-up codes emerged and were adapted based on the analysis of participants' think-aloud sessions, particularly their interactions with interface elements and their engagement with statistics. Codes were designed to be overlapping to capture the multifaceted nature of user experiences. To confirm reliability, the initial coding was reviewed at a later stage by the same researcher, and any deviations were corrected through iterative revisions. Frequency counts were kept for each code to weigh their impact on participant choices.
\section{Results}
This section is divided into 3 subsections. The first 2 subsections document feedback for Google Dataset Search and data.europe.eu, labeled '\textbf{P}' for positive and '\textbf{N}' for negative. Each of these subsections includes six user experience elements identified in the user task. All percentage values have been rounded to two decimal places.

While certain insights, such as interface design elements, are specific to each platform, most of the findings regarding mechanisms and functionalities can be applied more broadly to dataset search platforms. Therefore, the last subsection ~\ref{last sub} presents the development of 10 new design prototypes, based on observations of features common in dataset search platforms across various fields. 

\subsection{Google Dataset Search }
In the initial interaction, 68\% of participants found the search box, button, help, and feedback features accessible and user-friendly, describing the design as "neat and comfortable" (P1). 79\% appreciated the layout of the search results page, understanding the filters and praising their placement and the balance between preview and detailed information sections (P2). However, 26\% noted the language switch button was hard to find at the bottom left, and 16\% found its design confusing, suggesting the use of country flags or explicit labels (N1, N2). In multitasking scenarios, 21\% found the detailed dataset information compressed and unreadable (N3).

During the search process, 63\% of participants used system suggestions to refine keywords, improving result relevance (P7). Google Dataset Search effectively corrected misspellings but sometimes failed to retrieve datasets for certain terms, offering spelling checks instead (P8, P9). 21\% effectively used Boolean and Exact Phrase operators, although specific prefixes did not yield desired results (P10, N14, N15). 37\% abandoned datasets due to missing values (N16), and the system sometimes misinterpreted queries (N17). Different spellings and sentence-like queries affected result relevance, and the Exclude Words operator did not function as expected (N18-N21). Searching for help with operators was unproductive as the User Guide page was missing, causing confusion (N22).

In exploring datasets, participants focused on the title, update time, and dataset type, with a preference for content over the publisher (P13-P16). Ranking, especially in the top three listings, influenced choices. Participants valued descriptions, citation counts, and providers in the right-hand display section (P17). While 74\% were indifferent to license specifics, 68\% preferred its existence (P18). However, 53\% initially cared about ranking, though 26\% disregarded it and checked many datasets, sometimes scrolling to the page's bottom (N29). Incomplete titles, limited preview information, and dense descriptions were common frustrations (N30-N32). 37\% highlighted a lack of detailed file information (N33), and despite difficulties in packing more information into previews, 16\% still preferred a clean preview section (N34). After finding a suitable dataset, 21\% clicked on 'explore at' for more details, but 11\% found this process cumbersome due to switching between multiple tabs (N35).

Regarding filters and sorting, 68\% intuitively understood filter names and effects (P23). However, 32\% found it difficult to sort datasets by recency (N42). Multiple-choice filters like "topic" confused 26\% of participants due to the misleading "+1" display, with 16\% misinterpreting it as indicating only one topic selected (N43). 16\% preferred interactive animations for dropdown symbols, and 16\% found the broad time spans in the 'Last updated' filter problematic (N44, N45). 47\% were inconvenienced by filter resets upon search query modification (N46). Issues with datasets lacking usage permission tags and broad categorization within the 'Topics' filter were noted (N47, N48). 47\% expressed confusion about dataset classification after applying filters (N49), and 11\% ignored filters altogether due to their initial invisibility (N50).

In terms of dataset actions, all participants easily found and used the "Save," "Share," and "Cite" buttons (P25). However, 16\% preferred using their browser's built-in share function (N56). When asked about sharing methods, 37\% wouldn't use Facebook and Twitter options, preferring work-related contexts (N57). Additionally, 21\% found the dedicated email sharing feature less useful (N57). 21\% misunderstood the "Click to copy link" as informational text rather than an interactive button (N58), and 16\% noted that the website does not specify the current citation format and standard (N59).

Lastly, in feedback and help, 32\% appreciated explanations on dataset search and provision (P27). 84\% easily found the "Help" and "Feedback" buttons, with some browsing the "Help" page content independently (P28). However, 11\% encountered a '404 Page not found' error on the user guide page, increasing their reading time due to navigating multiple help pages (N62).

\subsection{data.europa.eu}
In the initial interaction with data.europa.eu, participants had mixed experiences. 11\% appreciated the comprehensive dataset summary and inclusion of datasets from 36 countries (P3). The page path display and window hover animations were positively noted by 26\% and 42\% of participants, respectively (P4, P5). However, 32\% of users on smaller laptop screens found the search box hard to spot (N7), and 53\% were distracted by unrelated windows and advertisements (N8). A UX designer highlighted the absence of search bar indicators (N9), and 37\% found filter selection cumbersome due to shifting positions (N10). Some participants (11\%) were frustrated by the inability to copy text from the preview section (N11), and color differentiation for tags was problematic for those with color blindness (N12). The 'show more' button was seen as redundant by 11\% of users (N13).

During the search process, 63\% found sorting by "Last modified" and "Last created" useful, especially under time constraints (P11). Separate search options for catalogues and datasets improved search efficiency for 21\% of participants (P12). However, 26\% noted that more search strings broadened the search scope instead of narrowing it (N23). Sorting by a single criterion led to extreme results for 16\% (N24), and the absence of transition animations for dropdown symbols was noted by the same percentage of users (N25).

In dataset exploration, 63\% of participants appreciated the use of color tags for dataset types (P20), and 42\% found the similar dataset feature beneficial (P21). The option to select 'Items per Page' was useful for 21\% (P22), although 47\% suggested using color tags for keyword matching instead (N36). Metadata like 'Update Time' and size were emphasized by 37\% as important for dataset assessment (N37). Confusion over non-English keywords and the 'Less Similar' tag was mentioned by 42\% and 16\% of participants, respectively (N39, N40). Some users found that datasets with the "less similarity" tag were not relevant (N41).

Regarding filters and sorting, 74\% praised the dual provision of filtering and sorting methods (P24). However, 58\% found page refreshes after each filter selection time-consuming (N51). This led to confusion for 37\%, as users thought filters were single-selection only (N52). The need for clearer indication of single-selection filters was noted by 16\% of participants (N53), while 11\% were confused by numbers in filter options' text (N54). Additionally, 26\% found the search box within each filter easy to overlook (N55).

For dataset actions, 42\% of participants found the citation function visible and comprehensive, offering multiple citation styles (P26). However, 11\% missed a direct LaTeX citation option (N60), and 58\% found comparing datasets cumbersome without a direct comparison feature on the website (N61).

Lastly, in feedback and help, 68\% of participants found the help documentation easy to locate and understand (P29). However, 47\% noted that while general information like "Contact Us" and "Documentation" was easy to find, options such as "Publications" and "Events" were often irrelevant when focused on searching for datasets (N63).

\subsection{New Prototypes for Dataset Search Platform}
\label{last sub}
\subsubsection{Initial Interaction} 
\begin{figure}[h]
    \centering
    \includegraphics[width=0.8\linewidth]{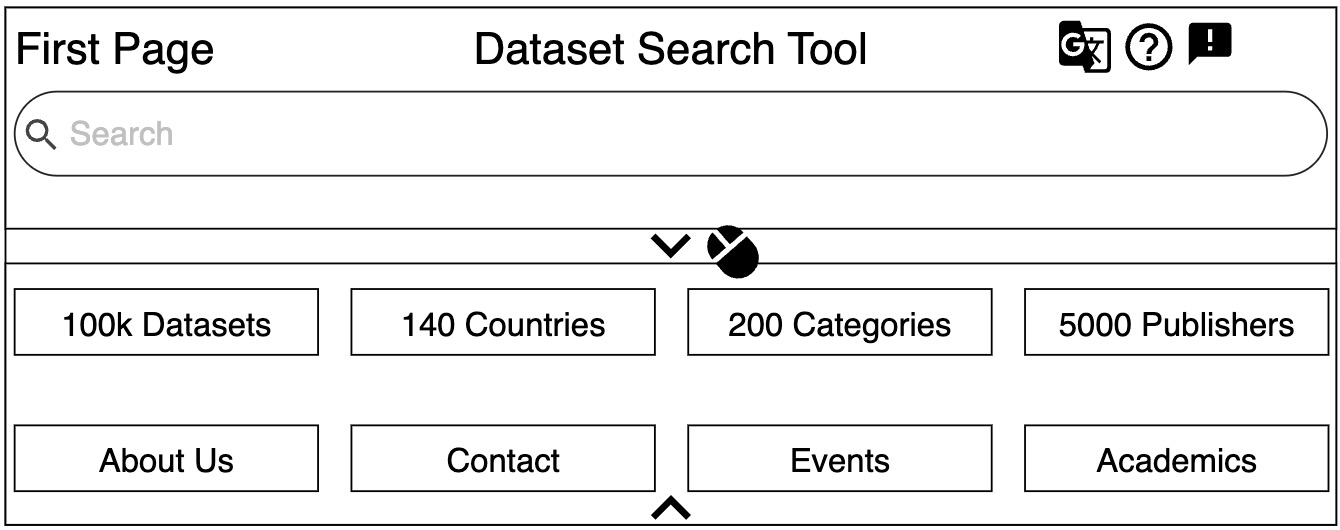}
    \caption{First Page of A Dataset Search Tool}
    \label{fig:p1}
\end{figure}

We suggest that first impressions are crucial for dataset search platforms. The search bar should be prominently displayed, and icons should be universally recognisable. A language switch button featuring familiar symbols, such as Google Translate, should be easily accessible. Secondary information, including website summaries and brand content, should either be collapsible or relocated to the bottom of the page. Visual cues such as breadcrumb navigation and subtle animations can further improve user engagement and navigation (Shown in Figure ~\ref{fig:p1}). These design elements should be consistently implemented across various screen sizes to ensure a user-friendly experience.

\subsubsection{Search Process} 
\begin{figure}[h]
    \centering
    \includegraphics[width=0.8\linewidth]{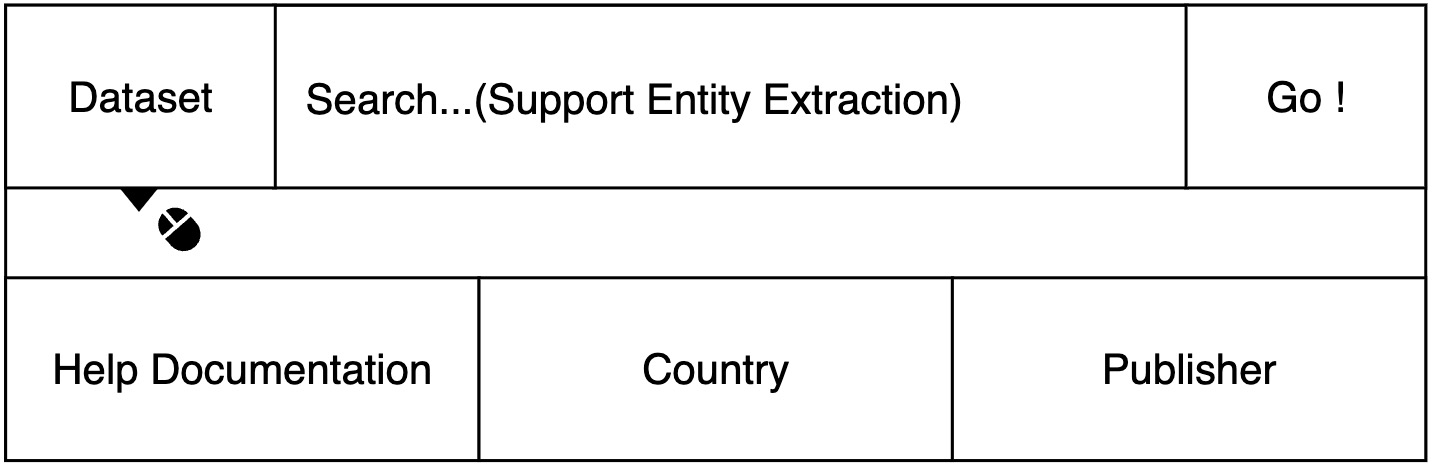}
    \caption{Search Bar Considering Varied User Needs}
    \label{fig:p2}
\end{figure}

\begin{figure}[h]
    \centering
    \includegraphics[width=0.8\linewidth]{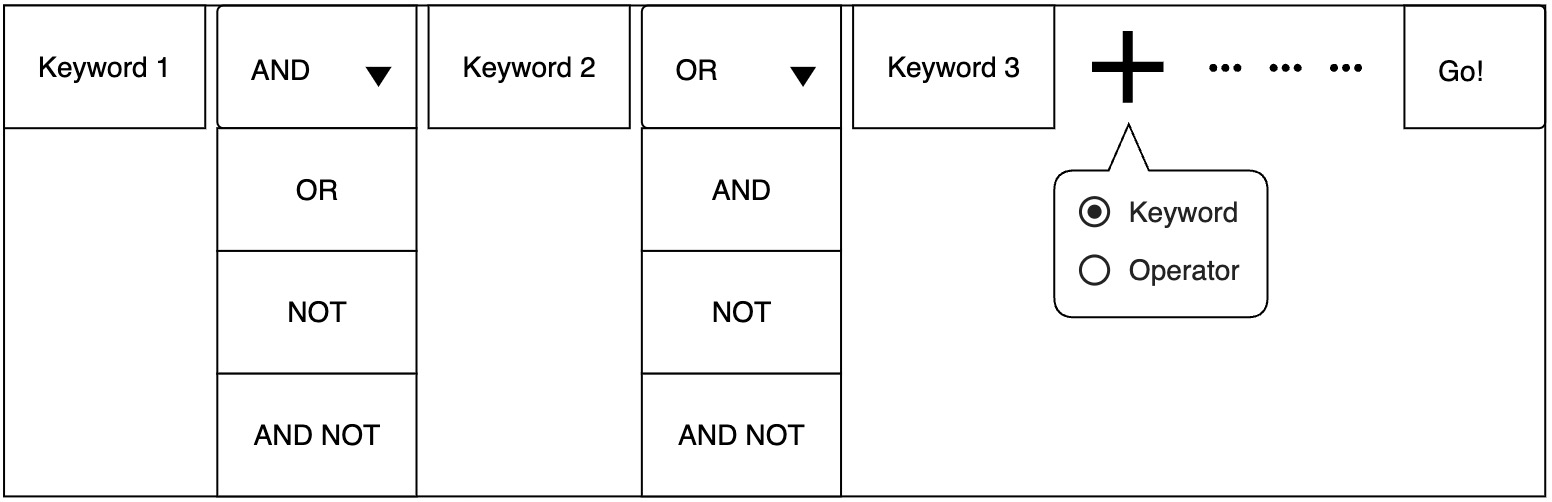}
    \caption{Search Bar Considering Enhanced Precision}
    \label{fig:p3}
\end{figure}

To enhance user interaction and platform efficiency, several improvements are recommended. Immediate auto-suggestions and spelling corrections could be provided through real-time prompts like "did you mean." Dataset search platform could also incorporate Natural Language Processing techniques such as entity extraction and synonym recognition for more accurate results. Special attention should be paid to varying user needs; for instance, new users might appreciate a help query option in the search bar, while other users may prioritize datasets based on factors like geographic location (Shown in Figure ~\ref{fig:p2}). Additionally, implementing common search operators like "AND," "OR," and "NOT" can refine search specificity. To make these operators more accessible, they could be visualized in the search bar interface, accompanied by dropdown menus or clickable icons. An 'add more' option could also be provided for scalable keyword and operator input. This approach not only informs users about advanced search tools but also enables them to generate highly targeted results from the beginning (Shown in Figure ~\ref{fig:p3}).

\subsubsection{Dataset Exploration}
\label{Dataset Exploration}
\begin{figure}[h]
    \centering
    \includegraphics[width=0.8\linewidth]{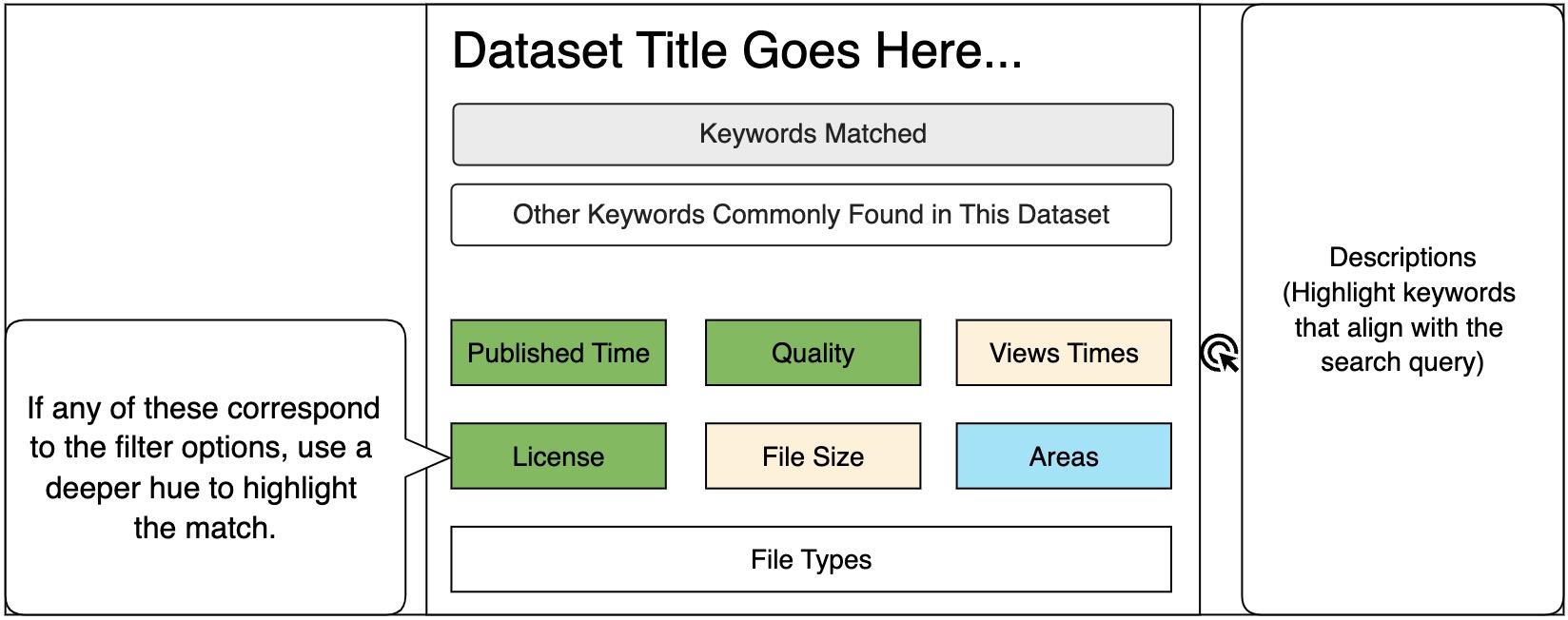}
    \caption{User-Focused Dataset Previews Section}
    \label{fig:p6}
\end{figure}

Dataset previews need to be brief but informative. They should underscore search terms, date of publication, quality of metadata, view counts, licence type, file dimensions, covered regions, and the nature of the dataset. Additionally, licence information should be streamlined to quickly signify if commercial use is permitted (See Figure ~\ref{fig:p6}). The previews should also be succinct while offering depth, including expandable segments for extra details. These segments can feature key terms, descriptions of columns, and ratios of missing values, presented in a bullet-point format.
\begin{figure}[h]
    \centering
    \includegraphics[width=0.8\linewidth]{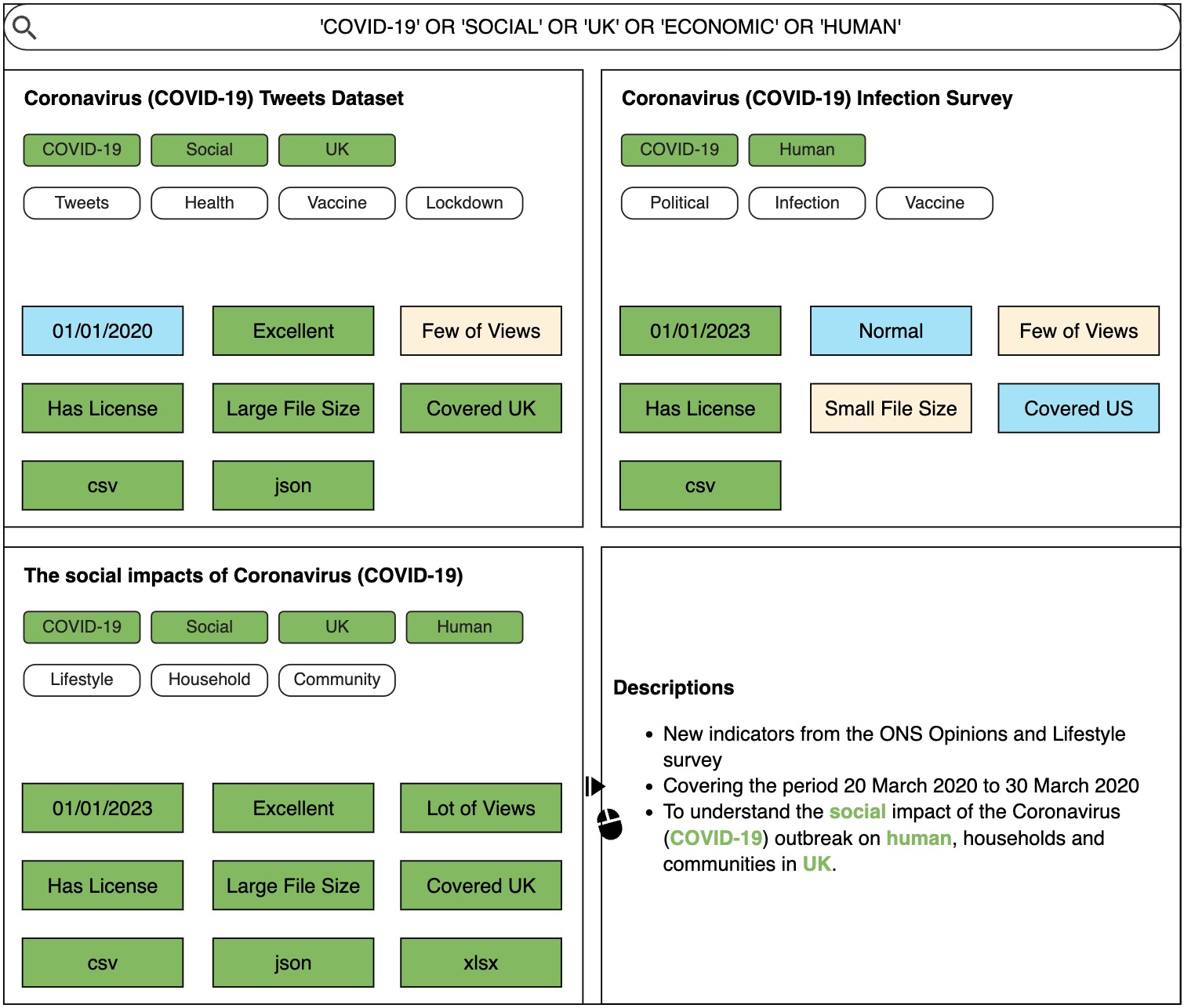}
    \caption{Simplified Result Display Section}
    \label{fig:p7}
\end{figure}

Given the numerous dataset previews, users can face reading fatigue. To counteract this, a colour-coded scheme could be employed to highlight datasets meeting user-defined criteria. For example, using dark green to signify criteria that match the user's definitions, intermediate blue for criteria that don't match, and pale orang for criteria that the user hasn't defined (Shown in Figure ~\ref{fig:p7}), while being accessible to colour-blind users, tested in Section ~\ref{color_test}.

\subsubsection{Usage of Filters and Sorting}
\label{Usage of Filters and Sorting}
\begin{figure}[h]
    \centering
    \includegraphics[width=0.8\linewidth]{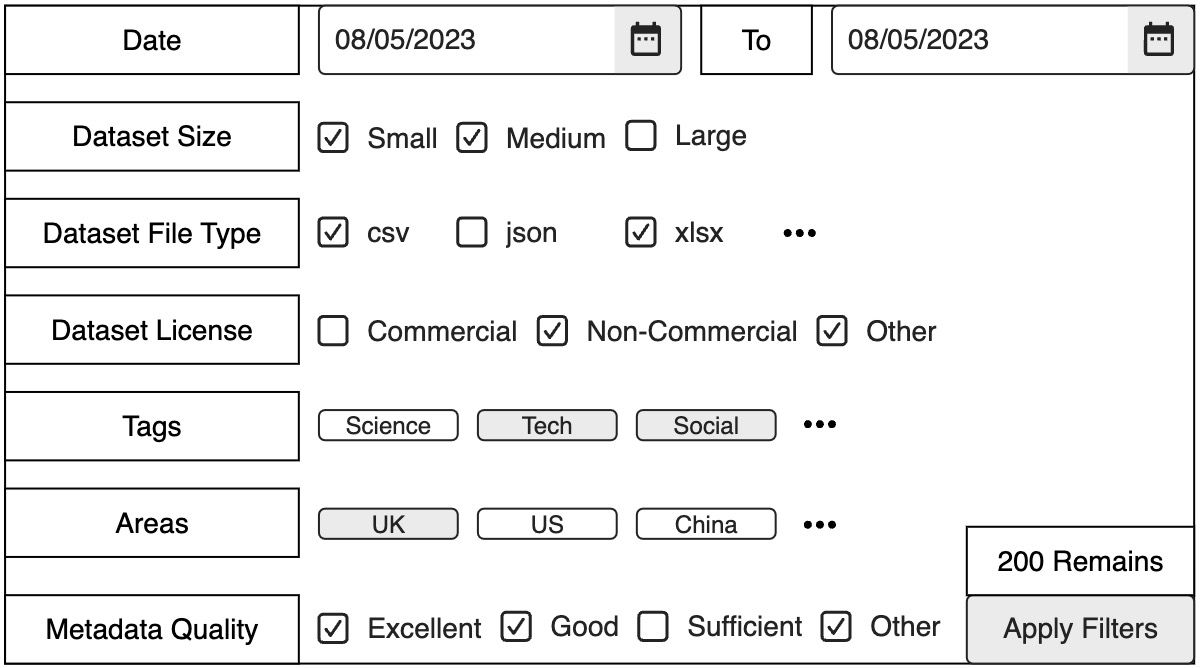}
    \caption{Improved Filter Design}
    \label{fig:p8}
\end{figure}

To enhance the user experience across platforms, several improvements are suggested. Filters should allow both single and multiple selections, symbolised by circles and boxes respectively. These filters should encompass a range of options including time range, dataset size, and license type. Ideally positioned at the top of the page, filters should not necessitate a full page refresh upon each selection. Instead, a subtle bottom-right corner notification should indicate the number of datasets that fit the selected criteria (Shown in Figure ~\ref{fig:p8}).
Sorting should employ dual criteria: initial sorting based on relevance, followed by secondary sorting according to publication date. These improvements aim for a more efficient and user-friendly dataset search process.

\subsubsection{Dataset Actions} 
\label{Dataset Actions}
\label{color_test}
\begin{figure}[h]
    \centering
    \includegraphics[width=0.8\linewidth]{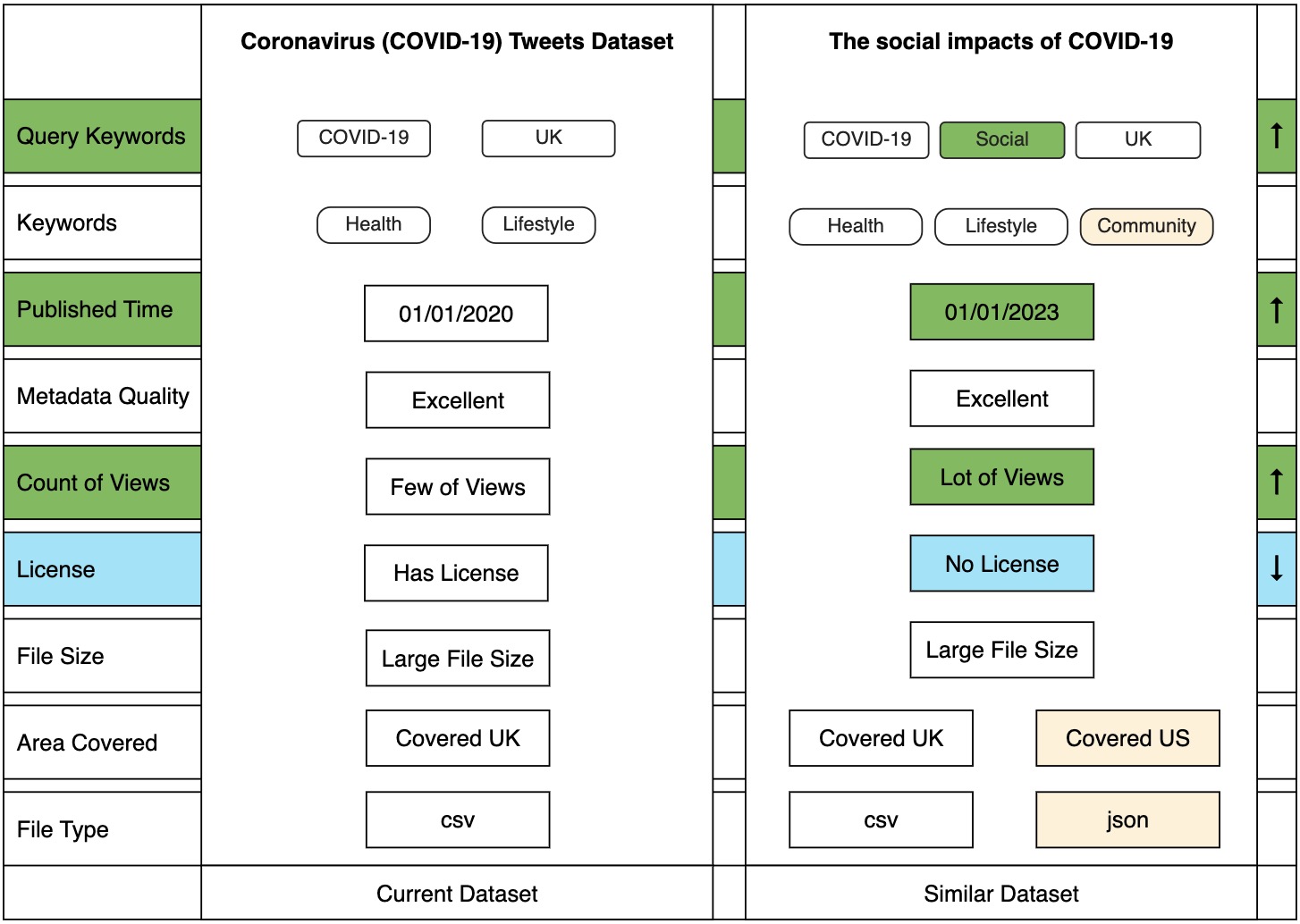}
    \caption{Datasets Comparison Page}
    \label{fig:p10}
\end{figure}

Improvements in dataset actions could be multifaceted. First, the citation functionalities across platforms should be expanded to include diverse styles like LaTeX, catering to academic requirements. Second, implementing a feature to allow users to directly compare datasets could be invaluable. The attributes of datasets should be colour-coded for more straightforward comparisons. A dark green fill could signify attributes where another dataset performs better than the currently selected one, a intermediate blue fill for areas where the current dataset excels, and a pale orange fill to highlight unique features not offered by the selected dataset. This visual guide would streamline the decision-making process, rendering it more intuitive and less cognitively taxing for users (Shown in Figure ~\ref{fig:p10}).

\begin{figure}[h]
    \centering
    \includegraphics[width=0.8\linewidth]{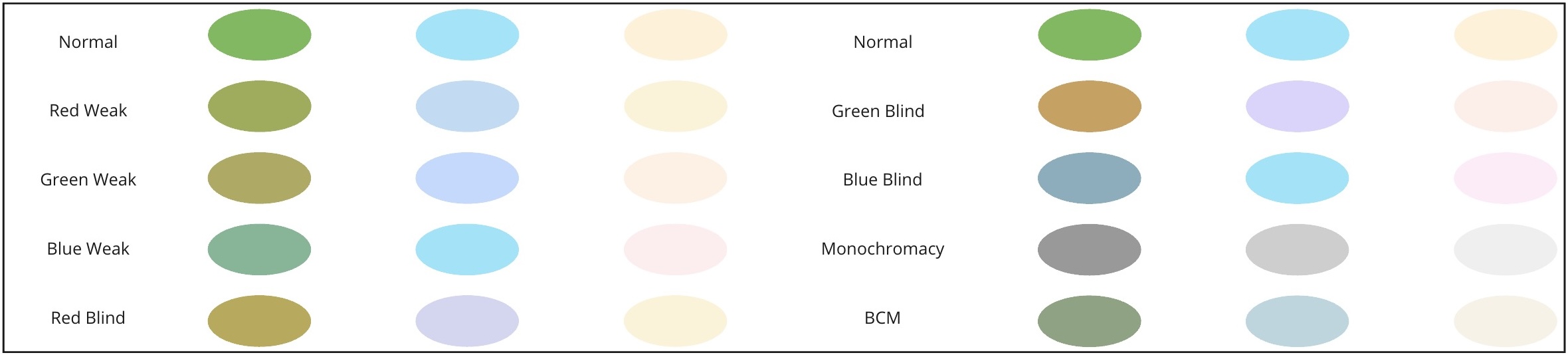}
    \caption{Eight Colour Blindness Tests}
    \label{fig:p11}
\end{figure}

The researcher used the Coblis Colour Blindness Simulator~\cite{colorblindness_2023} to test 8 types of colour blindness on the Figure ~\ref{fig:p10} provided . The results show clear variations in colour fills for each type of colour blindness, ensuring accessibility for those with colour vision deficiencies (Shown in Figure ~\ref{fig:p11}).

\subsubsection{Feedback and Help}
\begin{figure}[h]
    \centering
    \includegraphics[width=0.8\linewidth]{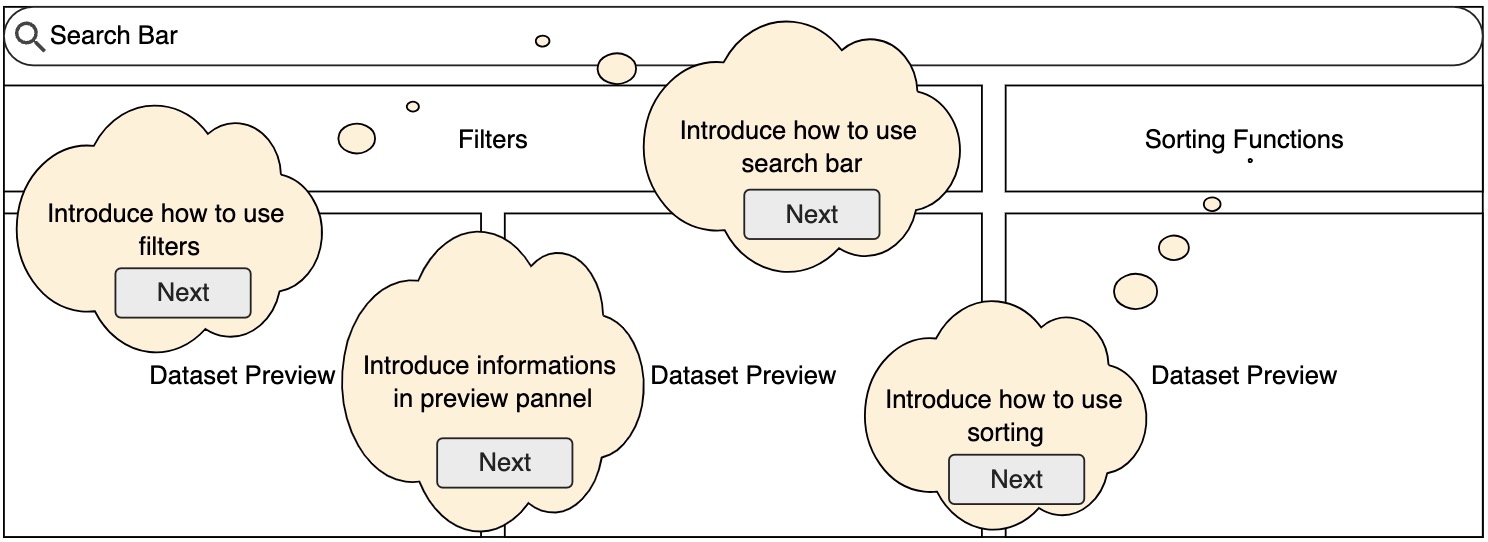}
    \caption{Tutorial Illustration}
    \label{fig:p12}
\end{figure}

For enhancements, both platforms could incorporate introductory tutorials to familiarize new users with their respective features. A step-by-step guide, necessitating users to acknowledge each tutorial segment before advancing, can ensure a more comprehensive understanding of the platform's functionalities. By instituting this level of guidance, platforms can increase user engagement, reduce initial learning curves, and subsequently, elevate the overall user experience (Shown in Figure ~\ref{fig:p12}).

\section{Discussion and Limitations}
Our study found that users experience significant reading fatigue due to the way metadata is presented on dataset search websites. There's a mismatch between the metadata displayed on search result previews and what users want to see, increasing the time taken for initial dataset selection. Detailed dataset pages are text-heavy and lack clear solutions for information presentation, with no quick dataset comparison feature. Current dataset websites also suffer from issues like excessive advertisements, irrelevant information, inadequate user interaction feedback, and animations. These platforms have adopted sorting and search methods from literature retrieval platforms, leading to inefficiency in dataset searches with extensive metadata.

To address these issues, our research proposes using distinct color fills to indicate metadata quality, reducing text descriptions to alleviate reading fatigue. Additionally, the color combinations are chosen to be friendly to color-blind users. We also simplify and intuitively display the most crucial metadata in search result previews, facilitating quick initial filtering for users, and an intuitive dataset comparison feature for efficient decision-making, using color fills to differentiate the quality of similar metadata between two datasets. We also suggest search function improvements, emphasizing the use of Boolean operators for effective metadata filtering, and website displays covering scope and functionality to save time in preliminary filtering. These changes aim to increase the efficiency of finding desired datasets. 

For limitations, such as user tasks set within a social science context, leading to search results predominantly related to social sciences. However, given the broad scope and abundance of social science datasets, they can be seen as representative of the current state of dataset metadata. Furthermore, since metadata is generally universal and our focus was on accessibility-related user experience, these aspects were generalized in our conclusions, minimizing their impact. Some of our findings, like the improvements for search bar and tutorials, resemble traditional literature search outcomes, but these were already existing issues with current dataset search platforms, they need to be acknowledged as well. Our primary focus was on exploring the accessibility of dataset metadata, specifically how to present it to improve search efficiency, precision, and reduce reading fatigue, addressing key user experience challenges today.

\section{Conclusion and Future Works}
This research assesses user interactions in dataset search on two platforms: Google Dataset Search and data.europa.eu. Think-aloud protocols and questionnaires are used as methods. Strengths and weaknesses in user experience are identified, particularly in terms of interface design and search functionality related to metadata accessibility. The developed prototypes focus on displays that highlight keywords and metadata, which are of primary interest to users. We also employ color schemes friendly to color-blind users to represent metadata quality, aiming to reduce reading fatigue and improve search efficiency. These features are especially evident in the detailed dataset pages, search results pages, and dataset comparison pages. This study contributes to the field of human-centered information retrieval. In future research, implementing eye-tracking will be a priority to gain more nuanced insights into user behaviour during dataset searches. The inability to employ GazeRecorder on platforms like Google Dataset Search and data.europa.eu suggests the need for alternative eye-tracking technologies. Physical eye-tracking devices could offer a solution to the limitations encountered with software-based approaches, as they are less likely to be hindered by website security protocols.


%
%
%
\bibliographystyle{splncs04}
\bibliography{samplepaper}

\end{document}